\documentstyle[nato,epsfig]{crckapb}

\newcommand{\be}{\begin{equation}}
\newcommand{\ee}{\end{equation}}
\def\aprle{\buildrel < \over {_{\sim}}}
\def\aprge{\buildrel > \over {_{\sim}}}
\begin{opening}

\title{NEWS ABOUT $\nu$'S}
\author{E.K. AKHMEDOV}

\institute{Centro de F\'\i sica das Interac\c c\~oes Fundamentais (CFIF)\\ 
Departamento de F\'\i sica, Instituto Superior T\'ecnico \\
Av. Rovisco Pais, P-1049-001 Lisboa, Portugal 
}

\end{opening}

\runningtitle{NEWS ABOUT $\nu$'S }

\begin{document}

\begin{abstract}
We review new results in neutrino physics, including the latest 
data of the Super-Kamiokande, SNO and K2K experiments.
\end{abstract}

\section{Introduction}

The recent evidence for neutrino oscillations \cite{SK1} provided us with 
the first firm evidence of physics beyond the standard model and opened a 
new an exciting era in neutrino studies. Neutrino physics is a very 
active branch of particle physics now, both experimentally and theoretically. 
The experimental data keep pouring in, and new experiments are either under 
way or in an advanced stage of planning. On the theoretical side, there are 
new analyses of the data, both in the 3-flavour and 4-flavour schemes; the 
analyses of the solar neutrino data are being extended to cover the ``dark 
side'' of the parameter space, not studied (or little studied) before; the 
reach of future planned experiments is being investigated and new experiments 
designed to eliminate the white spots on the map of neutrino properties are 
being suggested. In addition, since the discovery of neutrino oscillations a 
large number of models of neutrino mass was proposed and some old models were
reconsidered in the light of the new data. 

In the present lectures we review new results in neutrino physics, mainly
the experimental ones. We discuss the latest results of the solar and 
atmospheric neutrino experiments and their analyses; results from the reactor 
and accelerator experiments, including those from the first long-baseline 
accelerator neutrino experiment K2K, and then briefly discuss future
experiments and projects. Finally, we discuss how all the presently 
available data can be summarized concisely in terms of the
phenomenologically 
allowed structures of the neutrino mass matrix. 

We do not discuss neutrino mass model because of the lack of space; for 
recent reviews on this subject the reader is referred to 
\cite{Sm1,AltFer,Tan,Moh1,Fr,Barr1}.  

The data reviewed in the lectures given at the NATO ASI in Cascais included 
the results reported at the Neutrino 2000 Conference in Sudbury, June 16 - 21, 
2000. However, the present written version is updated to include the results 
reported up to November of 2000. 
For other recent reviews on neutrino physics see, e.g., \cite{Bil,Akh1}.

\section{General properties of neutrinos}
Direct kinematic searches of neutrino masses yield the following
upper limits \cite{Lob,Wein,PDG,Ron}:
\begin{eqnarray}
&& m_{\nu_1}< 2.5~\mbox{eV at 95\% c.l. (Troitsk)}\,; < 2.2~\mbox{eV at
95\% c.l. (Mainz)}\,;~~~~
\\ \label{m1}  
&& m_{\nu_2}< 170~\mbox{keV at 90\% c.l. (PSI}; ~\pi^+\to\mu^++\nu_\mu)\,;
~~~~\\ \label{m2}
&& m_{\nu_3}< 15.5~\mbox{MeV at 95\% c.l. (ALEPH,\,CLEO,\,OPAL}; ~\tau~
\mbox{decays)}\,.~~~~
\label{m3}   
\end{eqnarray}
Here $\nu_1$, $\nu_2$ and $\nu_3$ are assumed to be the primary mass 
components of $\nu_e$, $\nu_\mu$ and $\nu_\tau$, respectively. However,
since we know now that at least one mixing angle in the lepton sector is 
large, these limits may need a re-interpretation. In particular, the upper
bound on $m_{\nu_3}$ may in fact be much more stringent than the one in 
eq. (\ref{m3}). 

How many neutrino species are there? Electron and muon neutrinos have been 
known to exist since 1955 and 1964, respectively. Although there were 
strong theoretical and indirect experimental reasons to believe that 
there exist the third neutrino species, $\nu_\tau$, until recently it 
was not unambiguously experimentally detected. In July of 2000 the DONUT 
Collaboration reported the direct experimental evidence for $\nu_\tau$
\cite{DONUT}. 

Are there any more neutrino species? The number of the types of the standard 
neutrinos can be found from the $Z^0$ decay width. Indeed, neutrinos 
from the $Z^0$ decays are not detected, and therefore the difference
between the measured total width of the $Z^0$ boson and the sum of its
partial widths of decay into quarks and charged leptons, the so-called
invisible width, $\Gamma_{inv}=\Gamma_{tot}-\Gamma_{vis}=498 \pm 4.2$ MeV, 
should be due to the decay into $\nu\bar{\nu}$ pairs. Taking into
account that the partial width of $Z^0$ decay into one $\nu\bar{\nu}$ pair 
$\Gamma_{\nu\bar{\nu}}=166.9$ MeV one finds the number of the light active
neutrino species \cite{PDG}:
\be
N_\nu=\frac{\Gamma_{inv}}{\Gamma_{\nu\bar{\nu}}}=2.994 \pm 0.012\,,
\label{nnu}
\ee
in a very good agreement with the existence of the three neutrino flavours.
There are also indirect limits on the number of light ($m< 1$ MeV)
neutrino species (including possible electroweak singlet, i.e. ``sterile''
neutrinos $\nu_s$) coming from big bang nucleosynthesis. The number of
neutrino species in equilibrium with the rest of the universe at the 
nucleosynthesis epoch is 
\be
N_\nu < 3.3\,,
\ee
though this limit is less reliable than the laboratory one (\ref{nnu}),
and probably four neutrino species can still be tolerated \cite{sarkar}.
In view of (\ref{nnu}), the additional neutrino species, if exist, must be  
a sterile neutrino $\nu_s$. 

Are neutrinos Dirac or Majorana particles? The only practical way of 
answering this question, known at the moment, is to look for the neutrinoless 
double beta decay
\be
A(Z,N)\to A(Z\pm 2,N\mp 2)+2e^\mp \,.
\label{2beta0nu}
\ee
This process can only occur if neutrinos are Majorana particles. If 
$2\beta 0\nu$ decay is mediated by the standard weak interactions, the
amplitude of the process is proportional to the effective neutrino mass 
\be
A(2\beta 0\nu) \propto \sum_i U_{ei}^2 \, m_i\equiv
\langle m_{\nu_e}\rangle_{eff}\,,
\label{ampl}
\ee
where $U_{ai}$ is the lepton mixing matrix. Neutrinoless double beta decay 
was searched for experimentally but up to now have not been discovered.
The experiments allowed to put upper bounds on the effective Majorana
neutrino mass, the best limit coming from the Heidelberg -- Moscow experiment 
on $2\beta$ decay of $^{76}$Ge \cite{Klapdor}:
\be
\langle m_{\nu_e}\rangle_{eff}< 0.2 - 0.6 ~\mbox{eV}\,,
\label{meff}
\ee
depending on the value of the nuclear matrix element which is not
precisely known. If the $2\beta 0\nu$ decay is discovered, it will be
possible to infer the value of the effective Majorana neutrino mass
$\langle m_{\nu_e}\rangle_{eff}$. As follows from (\ref{ampl}), this would
give the lower limit on the mass of the heaviest neutrino.
For a recent review of $2\beta$ decay experiments, see \cite{Ejiri}.

\section{Atmospheric neutrinos
\label{atm}} 
are electron and muon neutrinos and their antineutrinos which are produced
in the hadronic showers induced by primary cosmic rays in the earth's 
atmosphere. The main mechanism of production of the atmospheric neutrinos
is given by the following chain of reactions:
\be
\begin{array}{lllll}
p(\alpha, ...)+Air&\rightarrow &\pi^{\pm}(K^{\pm})&+ & X  \\
&   &\pi^{\pm}(K^{\pm})&\rightarrow
&\mu^{\pm}+\nu_{\mu}(\bar{\nu}_{\mu})\\
& & & &\mu^{\pm}\rightarrow e^{\pm}+\nu_{e}(\bar{\nu}_{e})+
\bar{\nu}_{\mu}(\nu_{\mu})
\end{array}
\label{nuprod}
\ee
Atmospheric neutrinos can be observed directly in large mass underground
detectors predominantly by means of their charged current (CC) interactions:
\begin{eqnarray}
&  \nu_{e}(\bar{\nu}_{e})+A\rightarrow e^{-}(e^{+})+X\,,
\nonumber \\
&  \nu_{\mu}(\bar{\nu}_{\mu})+A\rightarrow \mu^{-}(\mu^{+})+X\,.
\label{nudet}
\end{eqnarray}
Calculations of the atmospheric neutrino fluxes predict the $\nu_\mu/\nu_e$ 
ratio that depends on neutrino energy and the zenith angle of neutrino 
trajectory, approaching 2 for low energy neutrinos and horizontal trajectories 
but exceeding this value for higher energy neutrinos and for trajectories
close to vertical.
The overall uncertainty of the calculated atmospheric neutrino fluxes is
rather large, and the total fluxes calculated by different authors differ
by as much as 20 -- 30\%. At the same time, the ratio of the muon to
electron neutrino fluxes is fairly insensitive to this uncertainty,  
and different calculations yield the ratios of muon-like to electron-like
contained events which agree to about 5\%. This ratio has been measured in
a number of experiments, and the Kamiokande and IMB Collaborations reported 
\mbox{smaller} than expected ratio in their contained events, with the
double ratio $R(\mu/e) \equiv [(\nu_\mu+\bar{\nu}_\mu)/
(\nu_e+\bar{\nu}_e]_{data}/[(\nu_\mu+\bar{\nu}_\mu)/(\nu_e+\bar{\nu}_e)]_{MC} 
\simeq 0.6$ where MC stands for Monte Carlo simulations. The discrepancy
between the observed and predicted atmospheric neutrino fluxes was called
the atmospheric neutrino anomaly. The existence of this anomaly was
subsequently confirmed by Soudan 2, MACRO and Super-Kamiokande experiments.
Most remarkably, the Super-Kamiokande (SK) Collaboration obtained a very
convincing evidence for the up-down asymmetry and zenith-angle dependent   
deficiency of the flux of muon neutrinos, which has been interpreted as an
evidence for neutrino oscillations. We shall now discuss the SK data and
their interpretation.

The SK detector is a 50 kt water Cherenkov detector (22.5 kt fiducial   
volume) which is overseen by more than 13,000 photomultiplier tubes. The 
charged leptons born in the CC interactions of neutrinos produce the rings
of the Cherenkov light in the detector which are observed by the phototubes.
Muons can be distinguished from electrons since their Cherenkov rings   
are sharp whereas those produced by electrons are diffuse. The SK 
Collaboration subdivided their atmospheric neutrino events into several
groups, depending on the energy of the charged leptons produced.
Fully contained (FC) events are those for which the neutrino
interaction vertex is located inside the detector and all final state
particles do not get out of it. FC events are further subdivided into
sub-GeV (visible energy $<1.33$ GeV) and multi-GeV (visible energy $>1.33$
GeV) events. Partially contained (PC) events are those for which the
produced muon
exits the inner detector volume (only muons are penetrating enough). The
average energy of a neutrino producing a PC event in SK is $\sim 15$ GeV.
Muon neutrinos can also be detected indirectly by observing the muons that
they have produced in the material surrounding the detector. To reduce the
background from atmospheric muons, only upward--going neutrino-induced
muons are usually considered. A rough estimate of the energy spectrum of the
upward--going muons has been obtained dividing them in two categories,
passing (or through-going) and stopping muons. The latter, which stop inside 
the detector, correspond to the average parent neutrino energy $\sim 10$ 
GeV, whereas for the through-going muons the average neutrino energy is
$\sim 100$ GeV.

The measurements of the double ratio $R(\mu/e)$ for contained events at SK
(1144 live days) give \cite{Sobel,Kaj}
\begin{eqnarray}
& & R=0.652\pm0.019\,(stat.)\pm 0.051\,(syst.)\quad \mbox{(sub-GeV)}\,,  
\nonumber \\
& & R=0.668\pm0.034\,(stat.)\pm 0.079\,(syst.)\quad \mbox{(multi-GeV)}\,.  
\label{R1}
\end{eqnarray} 
The value of $R$ for sub-GeV events is different from unity (to which it
should be equal in no-oscillation case) by more than $6\sigma$.

We shall now discuss the zenith angle distributions of the atmospheric
neutrino events. It should be remembered that the zenith angle distributions
of charged leptons which are experimentally measured do not coincide
with those of their parent neutrinos: for multi-GeV neutrinos the average
angle between the momenta of neutrinos and charged leptons is about 
$17^\circ$, whereas for sub-GeV neutrinos it is close to $60^\circ$. This
is properly taken into account in MC simulations. For PC events and upward 
going muons the correlation between the directions of momenta of muons and
parent neutrinos is much better. The distances $L$ traveled
by neutrinos before they reach the detector vary in a wide range: for
vertically downward going neutrinos (neutrino zenith angle $\Theta_\nu=0$)
$L\sim 15$ km; for horizontal neutrino trajectories ($\Theta_\nu=90^\circ$) 
$L\sim 500$ km; the vertically up-going neutrinos ($\Theta_\nu=180^\circ$)
cross the earth along its diameter and for them $L\sim 13,000$ km.

\begin{figure}[h]
\hbox to \hsize{\hfil\epsfxsize=10cm\epsfbox{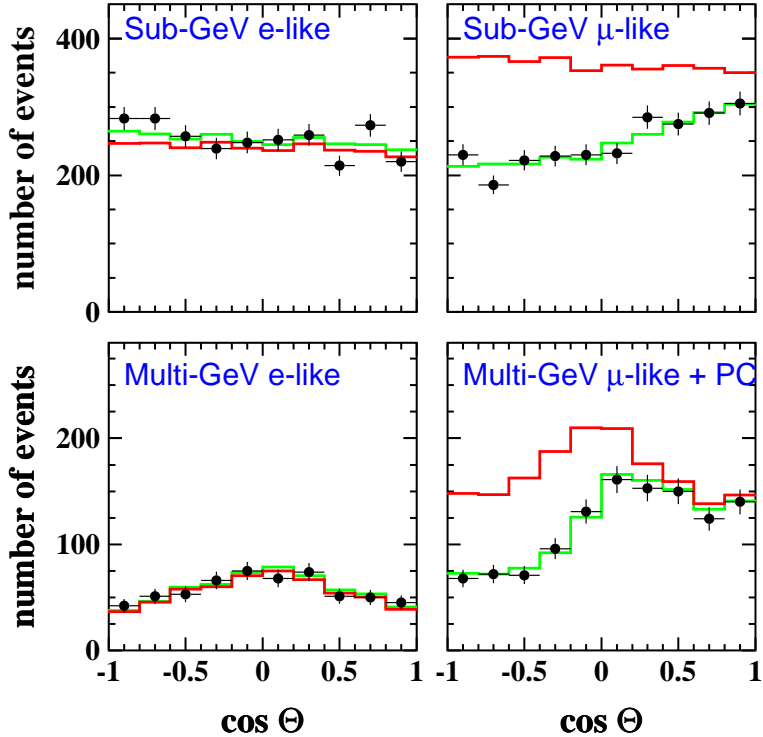}\hfil}
\caption{\small Zenith angle distributions for sub-GeV and
multi-GeV e-like and $\mu$-like events at SK (1144 live days). The 
dark-hatched lines show the (no-oscillations) Monte Carlo predictions; 
light-hatched lines show the predictions for $\nu_\mu \leftrightarrow
\nu_\tau$ oscillations with the best-fit parameters $\Delta m^2=3.2\times
10^{-3}$ eV$^2$, $\sin^2 2\theta=1.0$ [18]. }
\label{zen1}
\vspace{-0.5cm}
\end{figure}

In fig. \ref{zen1} the zenith angle distributions of the SK e-like and   
$\mu$-like events are shown separately for sub-GeV and multi-GeV contained
events. One can see that for e-like events, the measured zenith angle 
distributions \mbox{agree} very well with the MC predictions (shown by bars), 
both in the sub-GeV and multi-GeV samples, while for $\mu$-like events both
samples show zenith-angle dependent deficiency of event numbers compared
to expectations. The deficit of muon neutrinos is stronger for upward going 
neutrinos which have larger pathlengths. In the multi-GeV sample, there is
practically no deficit of events caused by muon neutrinos coming from the 
upper hemisphere ($\cos\Theta>0$), whereas in the sub-GeV sample, all 
$\mu$-like events exhibit a deficit which decreases with $\cos\Theta$. This 
pattern is perfectly \mbox{consistent} with oscillations $\nu_\mu 
\leftrightarrow \nu_\tau$ or $\nu_\mu \leftrightarrow \nu_{s}$ where
$\nu_s$ is a sterile 
neutrino. Muon neutrinos responsible for the multi-GeV sample are depleted
by the oscillations when their pathlength is large enough; the depletion
becomes less pronounced as the pathlength decreases ($\cos\Theta$ increases); 
for neutrinos coming from the upper hemisphere, the pathlengths are too short 
and there are practically no oscillations. Neutrinos responsible for the 
sub-GeV $\mu$-like events have smaller energies, and so their oscillation
lengths are smaller; therefore even neutrinos coming from the upper hemisphere 
experience sizeable depletion due to the oscillations. For up-going sub-GeV 
neutrinos the oscillation length is much smaller than the pathlength and they 
experience averaged oscillations. The solid line in fig. \ref{zen1} obtained 
with the $\nu_\mu \leftrightarrow \nu_\tau$ oscillation parameters in the
2-flavour scheme $\Delta m^2=3.2 \times 10^{-3}$ eV$^2$, $\sin^2 2\theta=1.0$ 
gives an excellent fit of the data.

\begin{figure}[htb]
\setlength{\unitlength}{1cm}
\hbox{\hfill
\epsfig{file=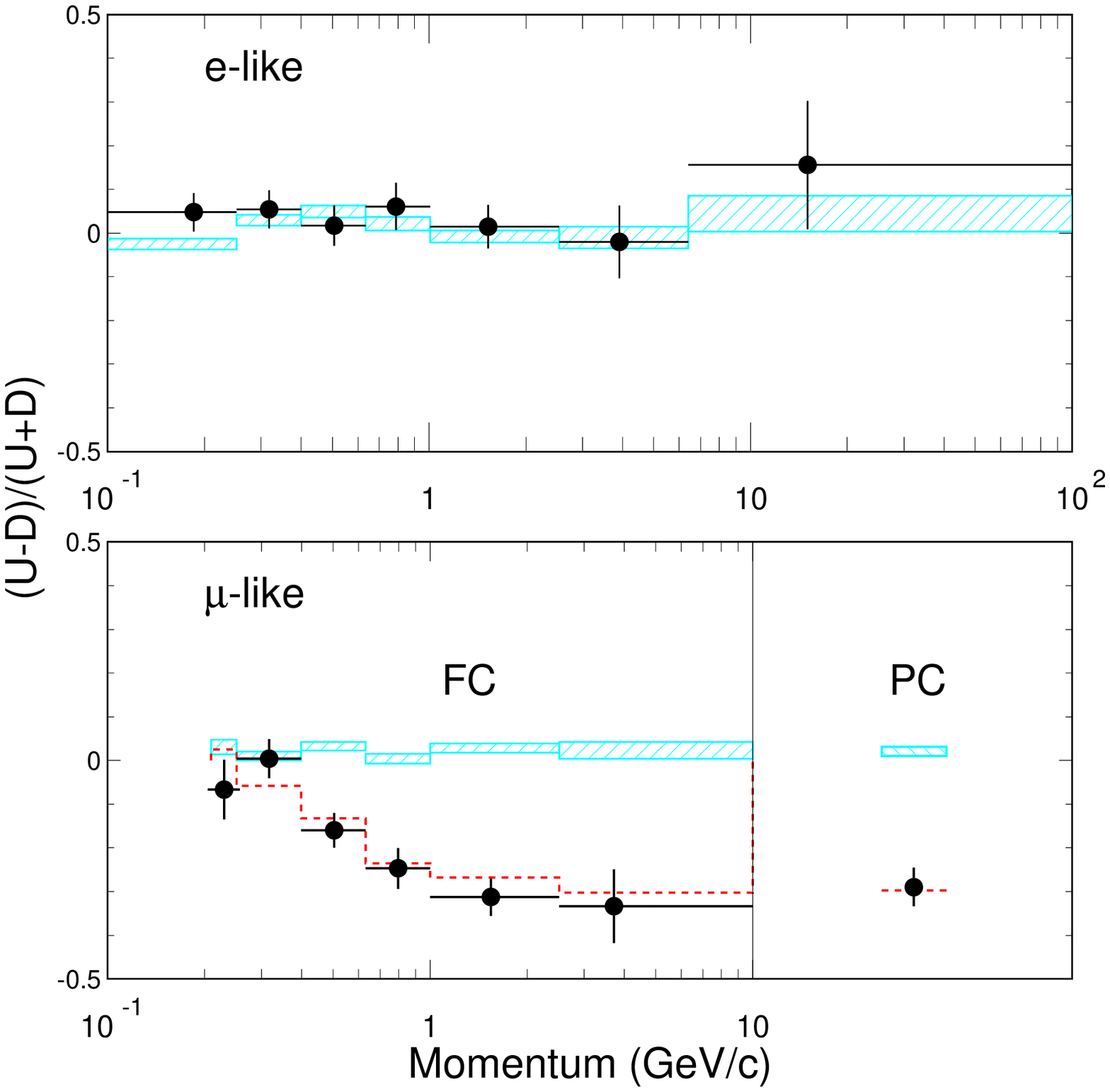,width=7.0cm}
\hspace{0.5cm}
\epsfig{file=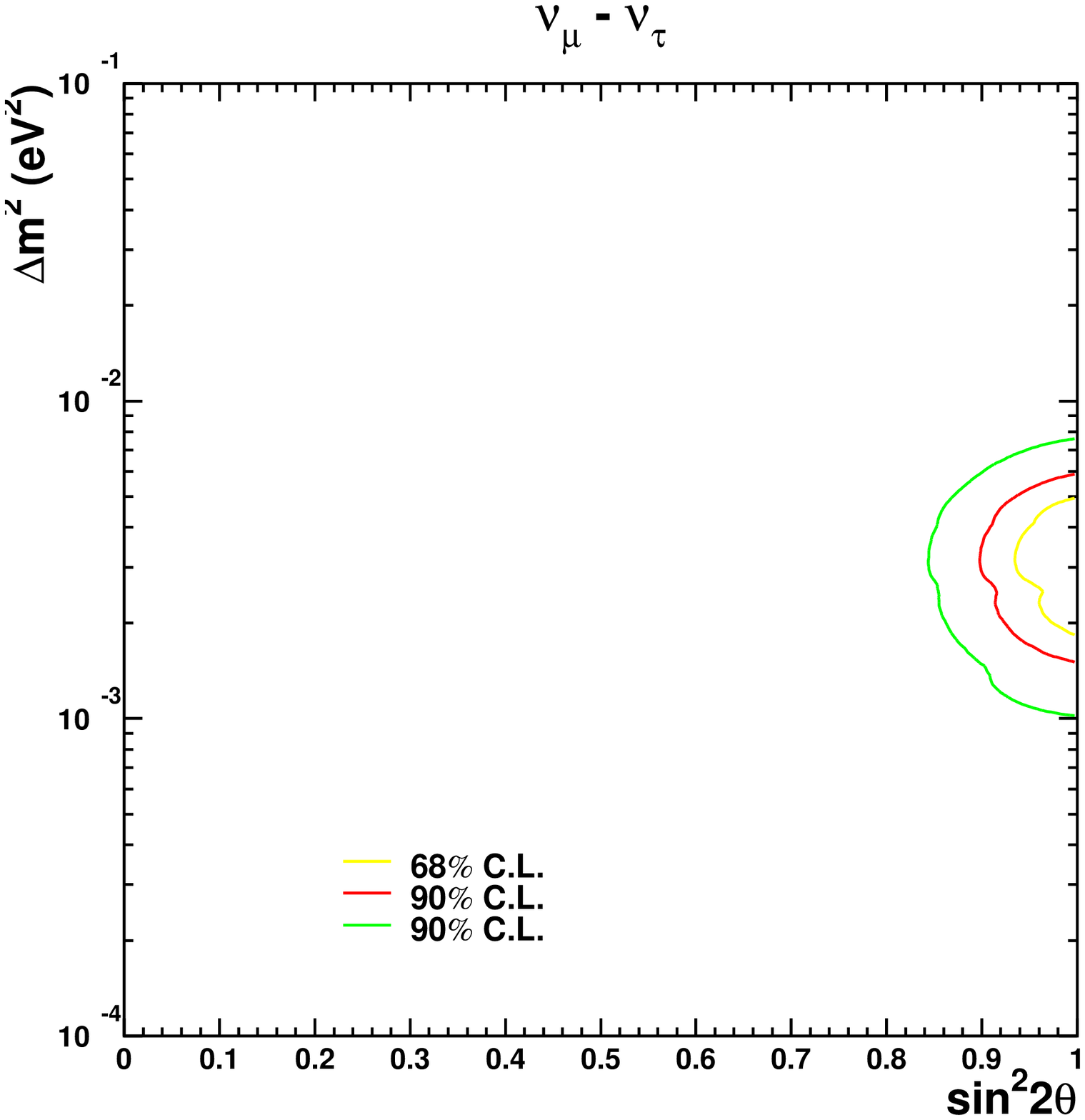,width=4.3cm,height=7.3cm}
\hfill}
\caption{\small Left panel: up-down asymmetry vs event momentum for single  
ring e-like and $\mu$-like events in SK (1144 live days). Hatched
bricks -- no oscillations, dashed line corresponds to $\nu_\mu
\leftrightarrow \nu_\tau$ oscillations with $\Delta m^2=3.2\times
10^{-3}$ eV$^2$, $\sin^2 2\theta=1$. Right panel: SK
allowed regions of oscillations parameters for $\nu_\mu \leftrightarrow 
\nu_\tau$ channel in 2-flavour scheme (FC + PC events) [18].} 
\label{asymm}
\vspace{-0.2cm}
\end{figure}

An informative parameter characterizing the distortion of the zenith angle
distribution is the up-down event ratio $U/D$, where up corresponds to the
events with $\cos\Theta <-0.2$ and down to those with $\cos\Theta > 0.2$.
The flux of atmospheric neutrinos is expected to be nearly up-down symmetric 
for neutrino energies $E\aprge 1$ GeV, with minor deviations coming from
geomagnetic effects which are well understood and can be accurately taken into 
account. In particular, at the geographical location of the SK detector
small upward asymmetry is expected, i.e. $U/D$ should be slightly bigger 
than 1. Any significant deviation of the up-down asymmetry of neutrino
induced events from the asymmetry due to the geomagnetic effects
is an indication of neutrino oscillations or some other new neutrino
physics. The $U/D$ ratio measured for the SK multi-GeV $\mu$-like events is 
\cite{Sobel,Kaj}
\be
U/D=0.54\pm 0.04\,(stat.) \pm 0.01\,(syst.)\,,
\label{A}
\ee
i.e. is below unity by about $9\sigma$! 
The dependence of the asymmetries for e-like FC and $\mu$-like FC+PC events 
on the event momentum is shown in fig. \ref{asymm} (left panel). One can
see that for e-like events the asymmetry $A\equiv (U-D)/(U+D)\simeq 0$ for 
all momenta. At the same time, for $\mu$-like events the asymmetry is close 
to zero at low momenta and decreases with momentum. This is easily 
understood in terms of the $\nu_\mu$ oscillations. For very small momenta, 
the oscillation length is small and both up-going and down-going neutrino
fluxes are depleted by oscillations to about the same extent; in addition, 
loose correlation between the directions of the momenta of the charged
lepton and of its parent neutrino tends to smear out the asymmetry at low
energies.
With increasing momentum the oscillation length increases, and the  
pathlength of down-going neutrinos becomes too small for oscillations to  
develop. 

The SK data show evidence for neutrino oscillations not only in their FC and 
PC $\mu$-like events: upward stopping and upward through-going events also
demonstrate zenith angle dependent deficiency of muon neutrinos consistent 
with neutrino oscillations, although the statistics for up-going muons is
lower than that for contained events. 
The combined analysis of the SK FC, PC and upward muon event data yields the 
best-fit values $\Delta m^2=3.2\times 10^{-3}$ eV$^2$ and  $\sin^2 2\theta=
1.0$, with the very high quality of the fit: $\chi^2/d.o.f.=135.4/152$. 
This value has to be compared with that of the no-oscillation hypothesis:
$\chi^2/d.o.f.=316.2/154$, which is a very poor fit. 

Are neutrino oscillations that are responsible for the depletion of the
$\nu_\mu$ flux $\nu_\mu\leftrightarrow \nu_\tau$ or $\nu_\mu\leftrightarrow
\nu_s$? For contained events, the oscillation probabilities in these two 
channels are nearly the same and the data can be fitted equally well in   
both cases, with very similar allowed ranges of the oscillation parameters. 
However, for higher energy PC and upward going events there are important
differences 
between  these two cases. In the 2-flavour scheme, $\nu_\mu\leftrightarrow
\nu_\tau$ oscillations are not affected by matter because the interactions
of $\nu_\mu$ and $\nu_\tau$ with matter are identical. However, sterile
neutrinos do not interact with matter at all, and therefore the $\nu_\mu 
\leftrightarrow \nu_s$ oscillations are affected by the matter-induced
potential $V_\mu-V_s=V_\mu$. At low energies, the kinetic energy difference 
$\Delta m^2/2E$ dominates over $V_\mu$, and the earth's matter effects are
unimportant. They become important at higher energies, when $\Delta
m^2/2E\sim V_\mu$; at very high energies, when $\Delta m^2/2E\ll V_\mu$,
matter strongly suppresses neutrino oscillations both in $\nu_\mu 
\leftrightarrow \nu_s$ and $\bar{\nu}_\mu\leftrightarrow \bar{\nu}_s$
channels. Therefore the oscillations of high energy neutrinos traveling 
significant distances in the earth should be strongly suppressed in this
case. Such a suppression was searched for in PC and upward trough-going
event samples, but has not been observed. This fact together with the 
analysis of the neutral current enriched multi-ring events allowed the SK
Collaboration to exclude pure $\nu_\mu\to \nu_s$ oscillations at the 99\%
c.l. \cite{Sobel,Kaj,SK2}. The oscillations into sterile neutrinos are, 
however, allowed in the 4-neutrino framework, with the weight that can
be as large as about 50\% \cite{Lisi1} 
\footnote{It should be noted, however, that the analysis of \cite{Lisi1}
does not include the SK neutral current enriched multi-ring event sample.}. 

Can $\nu_\mu\leftrightarrow \nu_e$ oscillations be responsible for the
observed anomalies in the atmospheric neutrino data? The answer is no, at
least not as the dominant channel. Explaining the data requires
oscillations with large mixing equal or close to the maximal one;
$\nu_\mu\leftrightarrow \nu_e$ oscillations would then certainly lead to
a significant distortion of the zenith angle distributions of the e-like
contained events, contrary to observations. In addition, for $\Delta m^2$
in the range $\sim 10^{-3}$ eV$^2$ which is required by the atmospheric
neutrino data, $\nu_\mu\leftrightarrow \nu_e$ oscillations are severely  
restricted by the CHOOZ reactor antineutrino experiment \cite{chooz},
which excludes these oscillations as the main channel of the atmospheric 
neutrino oscillations.

However, $\nu_\mu\leftrightarrow \nu_e$ and $\nu_e\leftrightarrow \nu_\tau$ 
can be present as subdominant channels of the oscillations of atmospheric
neutrinos. This may lead to interesting matter effects on oscillations of 
neutrinos crossing the earth on their way to the detector.
Matter can strongly affect $\nu_e\leftrightarrow \nu_{\mu,\tau}$ and
$\bar{\nu}_e\leftrightarrow \bar{\nu}_{\mu,\tau}$ oscillations leading to
an enhancement of the oscillation probabilities for neutrinos and
suppression for antineutrinos or vice versa, depending on the sign of
the corresponding mass squared difference. 

Since three neutrino flavours are known to exist, oscillations of atmospheric
neutrinos should in general be considered in the full 3-flavour framework  
(assuming that no sterile neutrinos take part in these oscillations). 
Three flavour analyses \cite{Sobel,Kaj,3fl,Concha} only slightly modify
the 2-flavour \mbox{results}, which is a consequence of the smallness of the
leptonic mixing parameter $U_{e3}$ 

Are the standard neutrino oscillations the sole possible explanation of  
the observed atmospheric neutrino anomalies? In principle, other explanations
are possible. Those include exotic types of neutrino oscillations  --
matter-induced oscillations due to flavour-changing interactions of neutrinos
with medium, oscillations due to small violations of the Lorentz or CPT
invariance or of the gravitational equivalence principle, and also neutrino 
decay. 
Exotic oscillations lead to periodic
variations of the $\nu_\mu$ survival probability with the oscillation lengths
$l_{osc}\propto E^{-n}$ where $n=0$ in the case of flavour-changing neutrino
interactions or violation of CPT invariance, and $n=1$ for oscillations
due to the violations of the Lorentz invariance or equivalence principle.
This has to be contrasted with $n=-1$ in the case of the standard neutrino   
oscillations. The energy dependence of the oscillation length can be 
tested in the atmospheric neutrino experiments as the energies of detected   
neutrinos span more than 3 orders of magnitude. The first analysis was 
performed in \cite{Lisi2}, and the authors found that the fit of the 
atmospheric neutrino data assuming oscillations with $l_{osc}\propto E^{-n}$ 
gives $n=-0.9\pm 0.4$ at 90\% c.l.. The SK Collaboration has recently 
performed a similar analysis of their full 1144 days sample of events,
which gave $n=-1.06\pm 0.14$ \cite{Sobel,Kaj}.  These results clearly 
favour the standard oscillations over the exotic ones. In contrast to this, 
the neutrino decay mechanism fits the SK data quite well, the quality of
the fit being as good as the one for the standard neutrino oscillations
\cite{Barger1}. The reason for this is that the oscillations and decay 
lead to very similar averaged survival probabilities for $\nu_\mu$. 
One could discriminate between the two solutions in the experiments like 
the recently proposed MONOLITH \cite{Mon}. The MONOLITH detector will have
a much better $L/E$ resolution than that of the SK, and therefore will be 
able to clearly detect the full first oscillation swing of the $\nu_\mu$
survival probability (i.e. not only disappearance but also reappearance of
$\nu_\mu$) in the case of the neutrino oscillation solution of the atmospheric 
neutrino anomaly.   

\section{The solar neutrino problem
\label{SNP}}

\begin{figure}[t]
\setlength{\unitlength}{1cm}
\begin{center}  
\vspace{-0.6cm} 
\epsfig{file=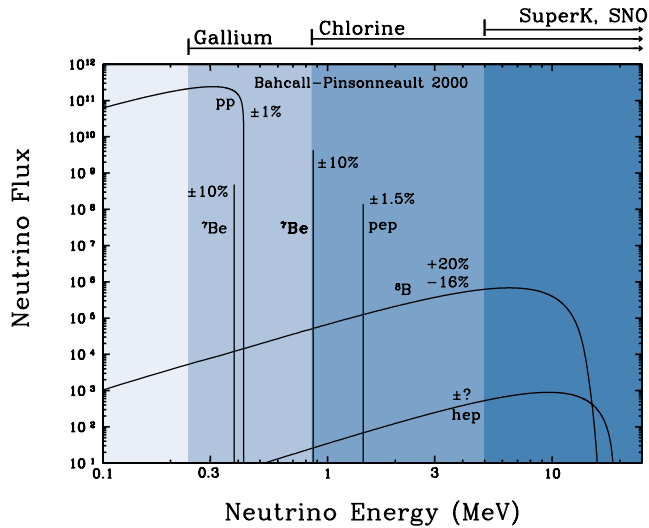,width=7.5cm,angle=270}
\end{center}
\vspace{-0.3cm} 
\caption{\label{spectrum}\small Solar neutrino spectrum and estimated 
theoretical errors of fluxes. The thresholds of solar neutrino experiments 
are indicated above the figure. {}From [28].}
\vspace{-0.3cm}
\end{figure}

The first experiment in which the solar neutrinos have been observed was
the Homestake experiment of Davis and his collaborators. It is based on the 
reaction
\be
\nu_e+{\rm ^{37}Cl}\to {\rm ^{37}Ar}+e^-\,.
\label{Cl}
\ee
The energy threshold of reaction (\ref{Cl}) is 0.814 MeV, so only the $^8$B 
and $^7$Be and $pep$ neutrinos are detected in the Homestake experiment (see 
fig.~\ref{spectrum}), the largest contribution coming from the $^8$B
neutrinos.  
The two gallium solar neutrino experiments, SAGE and Gallex, employ 
the reaction
\be
\nu_e+{\rm ^{71}Ga}\to {\rm ^{71}Ge}+e^-\,.
\label{Ga}
\ee
The energy threshold of this reaction is 0.233 MeV, and so
the gallium experiments can also detect the lowest energy $pp$ neutrinos.
The next two experiments -- Kamiokande and its up-scaled version
Super-Kamiokande -- are the water Cherenkov detectors and use the
neutrino-electron scattering reaction
\be
\nu_a+e^-\to \nu_a + e^-\,.
\label{nue}
\ee
to detect solar neutrinos.
This reaction has zero physical threshold, but one has to introduce energy
cuts to suppress the background. In the Kamiokande experiment solar neutrinos 
with the energies $E>7.5$ MeV were detected, whereas the threshold used by 
Super-Kamiokande (SK) in their analysis is at present 5.5 MeV. With these 
energy cuts, the Kamiokande and SK detection rates
are only sensitive to the $^8$B component of the solar neutrino flux 
(the highest-energy $hep$ neutrinos give a negligible contribution to the
total detection rates as their flux is very low).

In all five solar neutrino experiments (Homestake, Gallex, SAGE, Kamio-kande 
and SK) fewer neutrinos than expected were detected, the degree of deficiency 
being different in the experiments of different types (see table I).  
This has been called the solar neutrino problem. 

The solar neutrino problem is not just the problem of the deficit of the
observed neutrino flux: 
results of different experiments seem
to be inconsistent with each other. 
In the absence of new neutrino physics, the energy spectra of the various   
components of the solar neutrino flux are given by the standard nuclear
physics and well known, and only the total fluxes of these components may
be different from those predicted by the standard solar models. 
The fluxes inferred from different experiments are not consistent with
each other, and in fact the best fit value of the the $^7$Be neutrino flux 
is negative! One is then led to conclude that neutrinos are not standard. 


\begin{table}
\caption{Detection rates in five solar neutrino experiments. 
Units are SNU (1 SNU = $10^{-36}$ captures per target atom per second) for
all the experiments except Kamiokande and SK, for which  
they are $10^6 cm^{-2}s^{-1}$. From [30]. }
\begin{tabular}{lccc} 
\vspace*{-0.3cm}
\\ \hline \hline
Experiment &  Data & Theory (BP98) &   Data/Theory  \\
\hline
Homestake  &  $2.56\pm0.16\pm0.14$ & $7.7\pm^{1.2}_{1.0}$ & $0.33\pm0.027$ \\ 
Kamiokande    &  $2.80\pm0.19\pm 0.33$ & $5.15^{+1.0}_{-0.7}$ & 
$0.54\pm0.07$ \\
SAGE &  $75.4\,^{+7.0}_{-6.8}\,^{+3.5}_{-3.0}$ & $129^{+8}_{-6}$ &
$0.58\pm0.06$ \\
Gallex+GNO &  $74.1\,^{+6.7}_{-6.8}$ & $129^{+8}_{-6}$ &
$0.57\pm 0.06$ \\
Super-Kamiokande & $2.40\pm0.03\,^{+0.08}_{-0.07}$ & 
$5.15^{+1.0}_{-0.7}$&$0.465\pm0.016$ \\ \hline
\end{tabular}
\vspace*{-0.3cm}
\end{table}
There are several possible particle-physics solutions of the solar neutrino
problem, the most natural one being neutrino oscillations. 
The neutrino oscillation solution has become even more plausible after the  
strong evidence for atmospheric neutrino oscillations was reported by the   
SK Collaboration. Neutrino oscillations can convert a fraction 
of solar $\nu_e$ into $\nu_\mu$ or $\nu_\tau$ (or their combination).
Since the energy of solar neutrinos is smaller than the masses of muons 
and tauons, these $\nu_\mu$ or $\nu_\tau$ cannot be detected in the CC
reactions of the type (\ref{Cl}) or (\ref{Ga}) and therefore are invisible
in the chlorine and gallium experiments. They can scatter on electrons
through the neutral current (NC) interactions and therefore should contribute 
to the detection rates in water Cherenkov detectors. However, the cross
section of the NC channel of reaction (\ref{nue}) is about a factor of 6 
smaller than that of the CC channel, and so the deficit of the neutrino flux
observed in the Kamiokande and SK experiments can be explained. The 
probabilities of neutrino oscillations depend on neutrino energy, and the
distortion of the energy spectrum of the experimentally detected solar 
neutrinos, 
which is necessary to reconcile the data of different experiments, 
is readily obtained.

\begin{figure}[h]
\begin{center}
\leavevmode
\psfig{figure=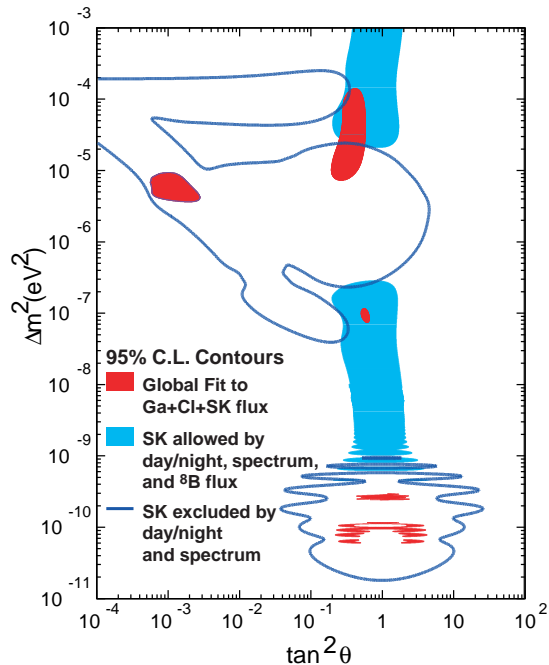,height=3.5in}
\end{center}
\caption{Solar neutrino parameter space: the dark areas show the global flux 
fit solutions.  The interiors of the dark lines indicate the SK excluded 
regions; the light shaded areas indicate the SK allowed regions [31].}  
\label{SKglobal}
\vspace*{-0.2cm}
\end{figure}   

The oscillations of solar neutrinos inside the sun can be strongly enhanced
due to the MSW effect \cite{MSW} and the solar data can be fitted even
with a very small vacuum mixing angle. Solar matter can also influence
neutrino oscillations if the vacuum mixing 
angle is not small. The allowed values of the neutrino oscillation parameters
$\tan^2\theta$ and  $\Delta m^2$ which fit the detection rates in the
chlorine, gallium and water Cherenkov experiments in the 2-flavour scheme are 
shown in fig. \ref{SKglobal} for 
oscillations of $\nu_e$ into active neutrinos (dark shaded areas). In the
case of the matter enhanced oscillations, there are three allowed ranges of 
the parameters corresponding to the small mixing angle (SMA), large mixing 
angle (LMA) and low $\Delta m^2$ (LOW) MSW solutions. 
There is also the vacuum oscillation (VO) solution corresponding to
very small values of $\Delta m^2$, for which the neutrino oscillation
length 
for typical solar neutrino energies ($\sim$ a few MeV) 
is comparable to the distance between the sun and the earth. This solution is 
also known as ``just so'' oscillation solution. The detection rates in the
five solar neutrino experiments can also be explained by $\nu_e\to\nu_{s}$
oscillations, for which there is only the SMA solution with the allowed 
region of parameters similar to that for oscillations into active neutrinos
\cite{Suzuki,recentfits}.

The use of the variable $\tan^2\theta$ instead of the usual $\sin^2 2\theta$ 
in fig. \ref{SKglobal} is worth a comment. The probability of 2-flavour 
neutrino oscillations in vacuum is invariant under the substitutions $\theta 
\to \pi/2-\theta$ or $\Delta m^2 \to - \Delta m^2$, but the oscillation 
probability in matter is not. It is, however, invariant under the combined 
action of these substitutions. To cover the full parameter space it is 
sufficient to assume $0\le \theta \le \pi/4$ and allow for both signs of 
$\Delta m^2$, or to assume that $\Delta m^2$ is always positive (which can
always be achieved by renaming the mass eigenstates $\nu_1\leftrightarrow 
\nu_2$) and let $\theta$ be in the full domain $[0,\pi/2]$. Usually, the 
first approach was adopted; however, the solutions of the solar neutrino
problem in the region $\Delta m^2<0$ have not been studied (except in  
the 3-neutrino \cite{Lisi3} and 4-neutrino \cite{Giu} frameworks). This
was motivated by the fact that there is no MSW enhancement for neutrinos in 
this region of parameters. However, in \cite{dGFM} it has been emphasized that 
if one allows for large enough confidence levels, or treats the solar $^8$B 
neutrino flux as a free parameter, or leaves the Homestake result out, 
solutions in this ``dark side'' of the parameter space exist, provided
that the mixing angle is close to the maximal one. It is convenient to assume 
$\Delta m^2>0$ and plot the allowed regions of the parameter space in the 
plane ($\tan^2\theta$, $\Delta m^2$) with $0\le \theta \le \pi/2$; in the 
conventional approach one would need two separate plots for $\Delta m^2>0$
and $\Delta m^2<0$. 

Solar neutrinos detected during night travel some distance inside the
earth on their way to the detector, and their oscillations can be affected  
by the matter of the earth. In particular, a fraction of $\nu_\mu$ or
$\nu_\tau$ produced as a result of the solar $\nu_e$ oscillations can   
be reconverted into $\nu_e$ by oscillations enhanced by the  matter of the
earth. The day/night difference due to the earth ``regeneration'' effect 
(and in general the zenith angle dependence of the neutrino signal) can in 
principle be observed in real-time experiments, such as SK. The day/night
effect is expected to be appreciable in the case of the LMA solution, but 
very small in the case of the SMA solution. For the LOW solution, the
day/night effect is expected to be quite sizeable in the low-energy part
of the solar neutrino spectrum (in particular, for $^7$Be neutrinos), but 
small for the high-energy part detected by SK.  
There is no day/night effect in the case of the VO solution. 
%

\begin{figure}[htb]
\begin{center}
\leavevmode
\psfig{figure=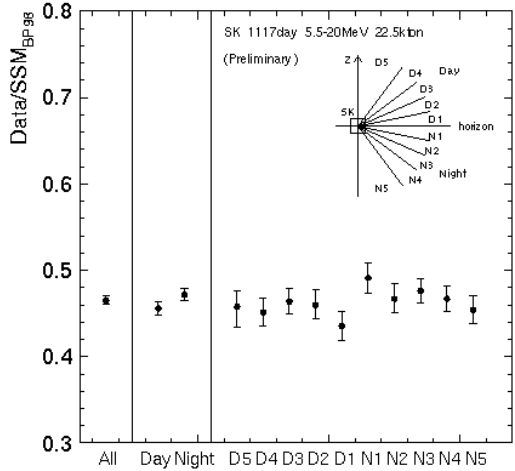,height=2.5in}
\end{center}
\caption{Zenith angle dependence of the solar neutrino flux measured by 
the Super-Kamiokande experiment [31].}
\label{daynight}
\end{figure}

The zenit angle dependence of the solar neutrino signal measured by the 
SK is shown in fig. \ref{daynight}. 
The value of day/night asymmetry measured in the SK experiment is 
\cite{Suzuki} 
\be
\frac{D-N}{(D+N)/2}=-0.034\pm 0.022\,(stat.)^{+0.013}_{-0.012}\,(syst.)\,,
\label{dn}
\ee
i.e. shows an excess of the night-time flux at about 1.3$\sigma$. 
This excess, however, is not statistically significant. 
The smallness of the SK day/night asymmetry results in the exclusion of 
the lower-$\Delta m^2$ part of the LMA allowed region (see fig. 4).    
This is a good news for future very long baseline accelerator experiments 
as it improves the prospects of observation of CP violation in neutrino
oscillations. 
The zenith angle event dependence measured by the SK shows a rather flat 
distribution of the excess of events over different night-time zenith
angle bins. This is rather typical for the LMA and LOW solutions
(although for LOW solution one can expect some excess of events in the
second  night bin $N2$ \cite{Sm2}), whereas for the SMA solution one
normally expects the excess (or deficiency) to be concentrated in the 
vertical upward bin with zenith angles $\theta$ in the range
$-1<\cos\theta <-0.8$.


\begin{figure}[htb]
\hbox to \hsize{\hfil\epsfxsize=8cm\epsfbox{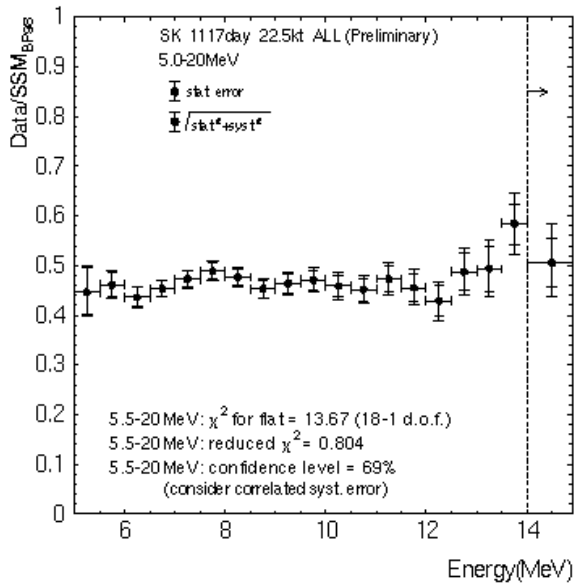}\hfil}
\caption{\small Recoil electron energy spectrum in the Super-Kamiokande
experiment normalized to the standard solar model prediction. From [31].}
\label{SKspectrum}
\end{figure}

Neutrino oscillations should result in
certain distortions of the spectrum of the detected solar neutrinos, which 
can be studied by measuring 
the recoil electron spectrum in the reaction (\ref{nue}). This spectrum has 
been measured in the SK experiment, and the results 
are shown in fig. (\ref{SKspectrum}). In the absence of the neutrino spectrum
distortion, the ratio measured/expected electron spectrum presented in
this figure should be a horizontal line. The measured relative spectrum 
is rather flat; in particular, the excess of high-energy events seen in the 
earlier SK data has essentially disappeared. This allowed SK to put an
upper limit on the flux of the 
$hep$ neutrinos: $f_{hep}<13.2 f_{hep}$(BP98), with the best fit value 
$f_{hep}=(5.4\pm4.5)f_{hep}$(BP98), where BP98 stands for the 
Bahcall-Pinsonneault 98 standard solar model. 

The flatness of the SK relative spectrum disfavours the VO solution and 
the larger-$\sin^2 2\theta$ part of the SMA parameter space; on the other
hand the small-$\sin^2 2\theta$ part of the SMA parameter space is disfavoured 
since it predicts a positive day-night asymmetry while the asymmetry measured 
by SK is negative (at $1.3\sigma$ level). This allowed the SK Collaboration to 
conclude that at present the VO and SMA solutions are disfavoured at 95\% 
c.l., while the LMA and LOW solutions are favoured (see fig. \ref{SKglobal}). 
Oscillations into sterile neutrinos are also disfavoured at 95\% c.l.. 

While this result is certainly interesting and shows the directions in 
which the solar neutrino data seem to lead us, one should clearly understand 
that in fact the VO and SMA solutions are {\em not} yet excluded. 
The solar neutrino data are not fully settled yet, and caution is advised 
when drawing conclusions. It is worth remembering that just a few years 
ago the LOW solution, which is now a perfectly respectable one, was 
disfavoured, and shortly before that it just did not exist. 
%
%
Obviously, more data are needed to clear the situation up. 

\begin{figure}[htb]
\hbox to \hsize{\hfil\epsfxsize=7.5cm\epsfbox{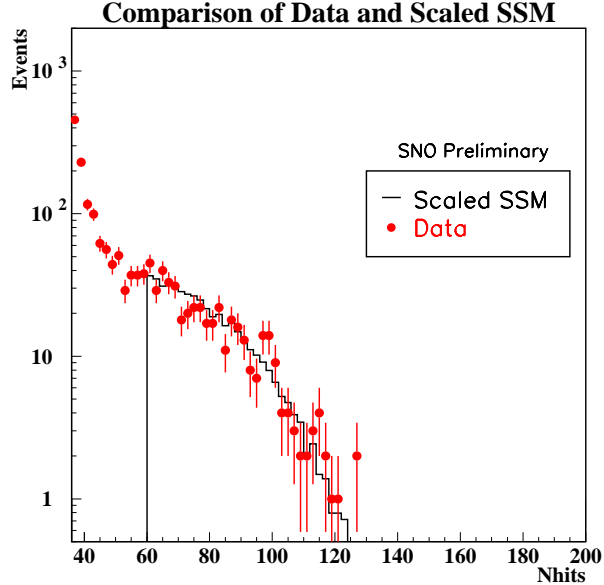}\hfil}
\vspace{0.4cm}
\caption{\small CC electron spectrum (distribution of events vs
number of PMT hits). The line is a scaled standard solar model prediction. 
The scaling factor is not specified. From [37]. }
\label{SNOspec}
\vspace{-0.2cm}
\end{figure}

Fortunately, several new experiments which can potentially resolve the 
problem are now under way or will be soon put into operation. The \mbox{SNO} 
(Sudbury Neutrino Observatory) experiment started taking data last year, and 
its first results were reported at the Neutrino 2000 Conference \cite{SNO}. 
The SNO detector consists of 1000 tons of heavy water, and is capable of
detecting solar neutrinos in three different channels:
\begin{eqnarray}
\nu_e+d \to p+p+e^- \quad\quad ({\rm CC}),~~E_{min}=1.44~{\rm MeV}\,, 
\label{CC}\\
\nu_a+d\to p+n+\nu_a \quad\quad ({\rm NC}),~~E_{min}=2.23~{\rm MeV}\,,   
\label{NC}
\end{eqnarray}
and $\nu_a e$ scattering process (\ref{nue}) which can proceed through
both CC and NC channels. The CC reaction (\ref{CC}) is very well suited
for measuring the solar neutrino spectrum: unlike in the case of $\nu_a e$  
scattering (\ref{nue}) in which the energy of incoming neutrino is shared
between two light particles in the final state, the final state of the
reaction (\ref{CC}) contains only one light particle -- electron, and a
heavy $2p$ system whose kinetic energy is relatively small. 
Therefore the electron energy is strongly correlated with the energy of 
the incoming neutrino. 
The CC electron spectrum measured by SNO (fig. \ref{SNOspec}) confirms 
the flat spectrum measured by SK. The absolute value of the solar neutrino 
flux measured in the CC reaction has not been given since the analysis of
the data is still under way. 

The cross section of the NC reaction (\ref{NC}) is the same for neutrinos
of all three flavours, and therefore oscillations between $\nu_e$ and 
$\nu_\mu$ or $\nu_\tau$ would not change the NC detection rate in the SNO
experiment. On the other hand, these oscillations would deplete the solar   
$\nu_e$ flux, reducing the CC event rate. Therefore the CC/NC ratio is
a sensitive probe of neutrino flavour oscillations. If solar $\nu_e$
oscillate into electroweak singlet (sterile) neutrinos, both CC and NC 
event rates will be suppressed.

The Borexino experiment is scheduled to start taking data in 2002. 
It will detect solar neutrinos through the $\nu_a e$ scattering with a
very low energy threshold, and will be able to detect the $^7$Be neutrino 
line. Different solutions of the solar neutrino problem
predict different degree of suppression of $^7$Be neutrinos, and their      
detection could help discriminate between these solutions. 
Observation of the $^7$Be neutrino line would be especially important
in the case of the VO solution. Due to the eccentricity of the earth's 
orbit the distance between the sun and the earth varies by about 3.5\%
during the year, and this should lead to varying oscillation phase (and
therefore varying solar neutrino signal) in the case of vacuum neutrino
oscillations. This seasonal variation can in principle be separated from 
the trivial 7\% variation due to the $1/L^2$ law which is not related to
neutrino oscillations 
\footnote{The seasonal dependence of the SK detection rate is in a good
agreement with the $1/L^2$ law, with $\chi^2/d.o.f.=4.1/7$ (goodness of
fit 76\%).}. 
Since the oscillation phase depends on neutrino energy, 
integration over significant energy intervals may make it difficult to 
observe the seasonal variations of the solar neutrino flux due to VO. The
$^7$Be neutrinos are monochromatic, which should facilitate the
observation of the seasonal variations at Borexino. 

Borexino will also be capable of confirming or refuting the LOW solution:
a strong day/night effect predicted for $^7$Be neutrinos by this solution  
should be clearly detectable at Borexino \cite{Murayama}.
On the other hand, the LMA solution will be tested by the KamLAND experiment, 
which will start taking data in 2001.  Although it will be a reactor neutrino 
experiment, its very long baseline will enable it to probe very small values 
of $\Delta m^2$, relevant for the LMA solution (see the next section).
One can hope that the combined data of the currently operating and
forthcoming experiments will allow to finally resolve the solar neutrino
problem. 

\section{Reactor and accelerator neutrino experiments
\label{reacc}}

In reactor neutrino experiments oscillations of electron antineutrinos into 
another neutrino species are searched for by studying possible depletion of 
the $\bar{\nu}_e$ flux beyond the usual geometrical one. 
These are the disappearance experiments, because the energies of the
reactor $\bar{\nu}_e$'s ($\langle E\rangle \simeq$ 3 MeV) are too small to
allow the detection of muon or tauon antineutrinos in CC experiments. Small 
$\bar{\nu}_e$ energy makes the reactor neutrino experiments sensitive to
oscillations with rather small values of $\Delta m^2$. 

Up to now, no evidence for neutrino oscillations has been
found in the reactor neutrino experiments, which allowed to exclude
certain regions in the neutrino parameter space. The best
constraints were obtained by the CHOOZ experiment in France \cite{chooz}. 
For the values 
of $\Delta m_{31}^2 \equiv \Delta m_{atm}^2$ in the SK allowed region 
$(1.5 - 5)\times 10^{-3}$ eV$^2$, the CHOOZ results give the following
constraint on the element $U_{e3}$ of the lepton mixing matrix: 
$|U_{e3}|^2 (1-|U_{e3}|^2)<0.055 - 0.015$ at 90\% c.l., i.e. $|U_{e3}|$ is
either small or close to unity. The latter possibility is excluded by solar 
and atmospheric neutrino observations, and one finally obtains
\footnote{
We use the parametrization of the $3\times 3$ lepton mixing matrix 
which coincides with the standard parametrization of the quark mixing
matrix.  
Notice that the fact that the latest SK data yield the allowed 
values of $\Delta m_{atm}^2$ which are somewhat lower than the previous
ones leads to slightly higher than before allowed values of $|U_{e3}|^2$.}
\be
\sin^2 \theta_{13}\equiv |U_{e3}|^2 \le 
(0.06 - 0.018)\, 
\quad
{\rm for}\quad\Delta m_{31}^2
=(1.5 - 5)\times 10^{-3}~{\rm eV}^2\,.
\label{choozres}
\ee
This is the most stringent constraint on $|U_{e3}|$ to date.

\begin{figure}[h]
\setlength{\unitlength}{1cm}
\hbox{\hfill
\epsfig{file=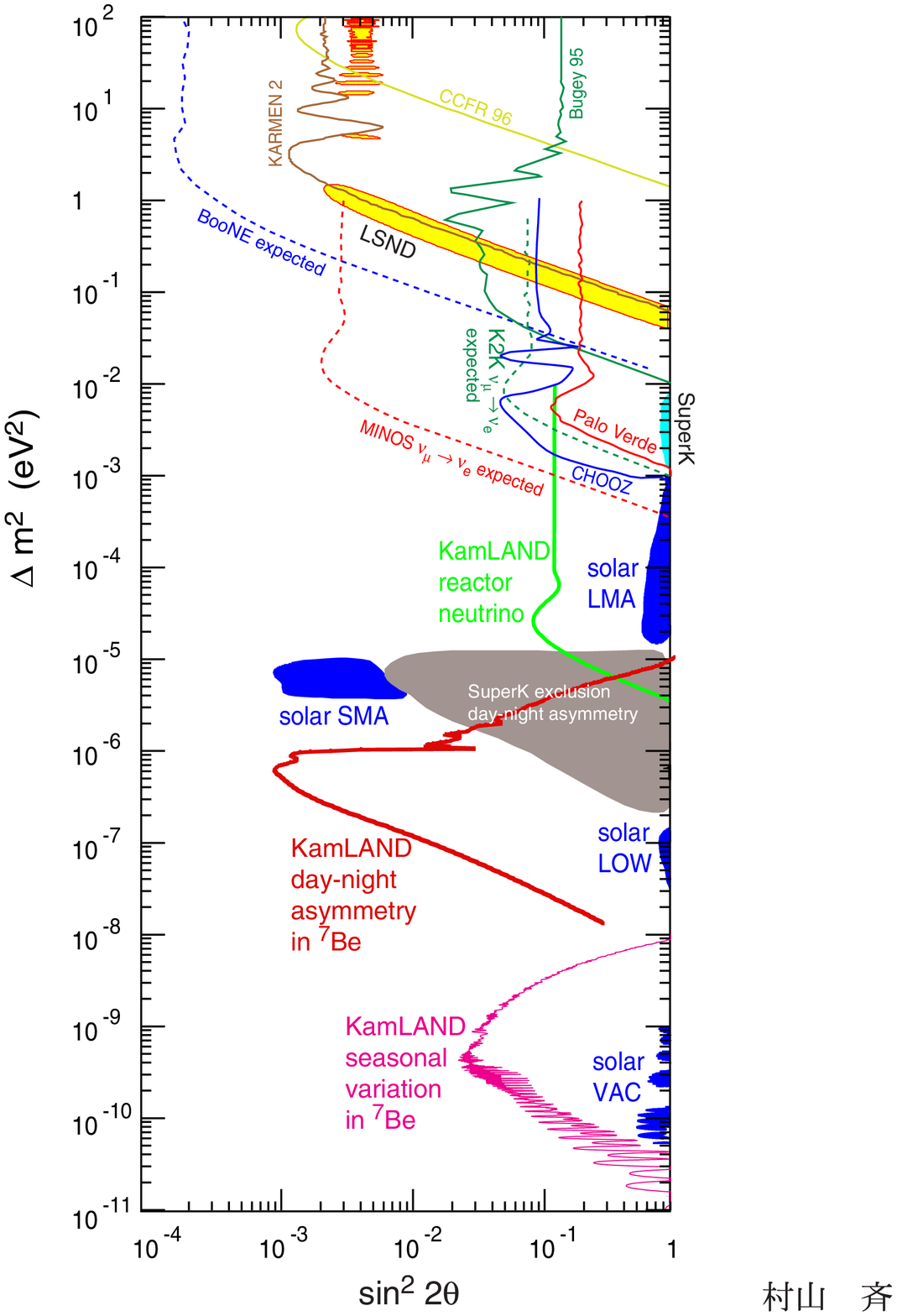,width=8.5cm,height=8.5cm}
\hspace{-1.8cm}
\epsfig{file=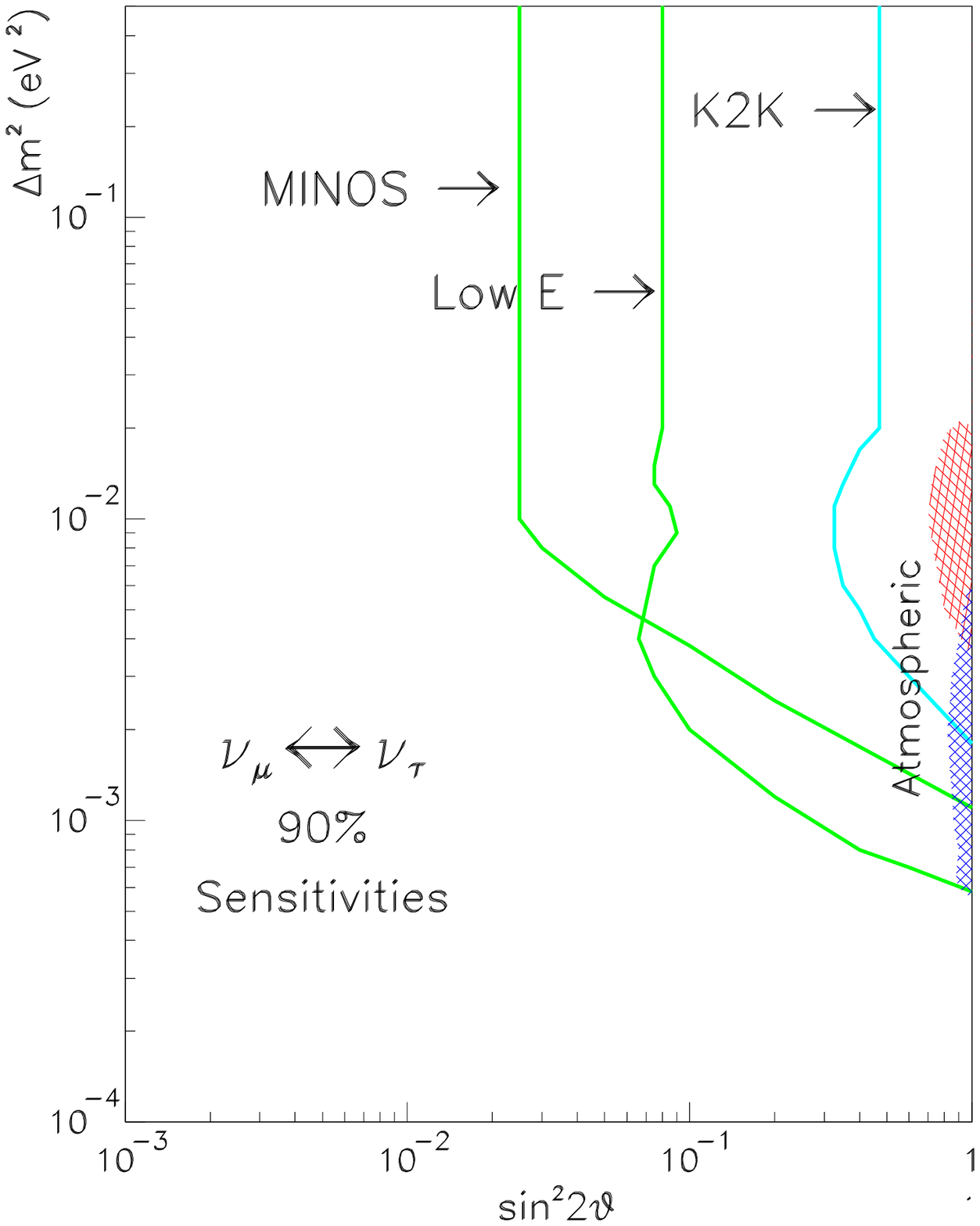,width=5.3cm,height=8.8cm}
\hfill}
\vspace{-0.2cm}
\caption{\small Left panel: results of the present and sensitivities of
future $\nu_e\leftrightarrow \nu_\mu$ oscillations searches (90\% c.l.). 
The plot by H. Murayama. Right panel: sensitivities of MINOS and K2K long
baseline experiments [41]. }
\label{future}
\vspace{-0.4cm}
\end{figure}

Presently, a long baseline reactor experiment KamLAND is under
construction in Japan. This will be a large liquid scintillator detector
experiment using the former Kamiokande site. KamLAND will detect electron
antineutrinos coming from several Japanese and Korean power reactors at an
average distance of about 180 km. KamLAND is scheduled to start taking data 
in 2001 and will be sensitive to values of $\Delta m^2$ as low as $4\times
10^{-6}$ eV$^2$, i.e. in the range relevant for the solar neutrino
oscillations! It is expected to be able to probe the LMA solution of the
solar neutrino problem (see fig. \ref{future}). It may also be able to
directly detect solar $^8$B and $^7$Be neutrinos after its liquid scintillator 
has been purified to ultra high purity level by recirculating through
purification. 


\begin{figure}[htbp]
\begin{center}
\leavevmode
\psfig{figure=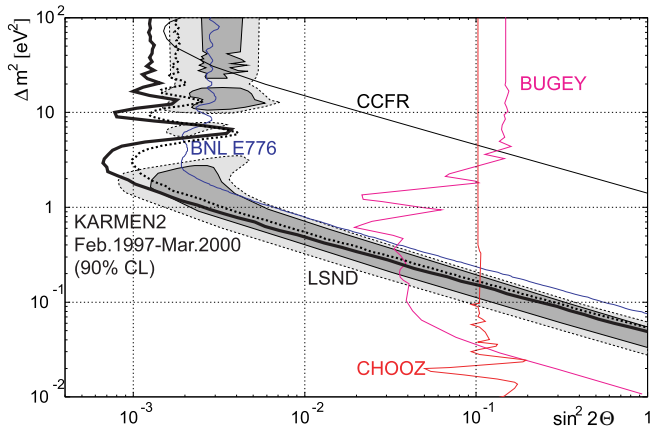,
bbllx=2pt,bblly=183pt,bburx=607pt,bbury=598pt,
height=2.3in}
\end{center}
\vspace*{-0.6cm}
\caption{LSND allowed parameter region for $\bar{\nu}_\mu\to \bar{\nu}_e$
oscillations (shaded areas) along with KARMEN, BNL E776, CCFR, CHOOZ and
Bugey exclusion regions [40]. }
\label{KARMLSND}
\vspace{-0.2cm}
\end{figure}

There have been a number of accelerator experiments looking for neutrino
oscillations. In all but one no evidence for oscillations was found and
constraints on oscillation parameters were obtained.  
The LSND Collaboration have obtained an evidence for $\bar{\nu}_\mu \to
\bar{\nu}_e$ and $\nu_\mu \to \nu_e$ oscillations \cite{LSND}. The LSND 
result is the only indication for neutrino oscillations that is a signal
and not a deficiency.
The KARMEN experiment \cite{Eitel} is looking for neutrino oscillations in
$\bar{\nu}_\mu \to \bar{\nu}_e$ channel. No evidence for oscillations has
been obtained, and part of the LSND allowed region has been excluded. In
fig. \ref{KARMLSND} the results from LSND and KARMEN experiments are shown
along with the relevant constraints from the BNL E776, CCFR, CHOOZ and
Bugey experiments. One can see that the only domain of the LSND allowed 
region which is presently not excluded is a narrow strip with $\sin^2
2\theta \simeq 1\times 10^{-3} - 4\times 10^{-2}$ and $\Delta m^2 \simeq
0.2 - 2$ eV$^2$. 

The existing neutrino anomalies (solar neutrino problem, atmospheric neutrino 
anomaly and the LSND result), if all interpreted in terms of neutrino 
oscillations, require three different scales of mass squared differences: 
$\Delta m^2_\odot \aprle 10^{-4}$ eV$^2$, $\Delta m^2_{atm} \sim 10^{-3}$ 
eV$^2$ and $\Delta m^2_{\rm LSND}\aprge 0.2$ eV$^2$. This is only possible
with four (or more) light neutrino species. The fourth light neutrino cannot
be just the 4th generation neutrino similar to $\nu_e$, $\nu_\mu$ and 
$\nu_\tau$ because this would be in conflict with the experimentally 
measured width of $Z^0$ boson [see eq. (\ref{nnu})]. It can only be an 
electroweak singlet (sterile) neutrino. Therefore the LSND result, if
correct, would imply the existence of a light sterile neutrino.

Out of all experimental evidences for neutrino oscillations, the LSND
result is the only one that has not yet been confirmed by other experiments. 
It is therefore very important to have it independently checked. This will
be done by the MiniBooNE (first phase of BooNE) experiment at Fermilab 
\cite{Bazarko}.   
MiniBooNE will be capable of observing both $\nu_\mu \to \nu_e$ appearance
and $\nu_\mu$ disappearance. If the LSND signal is due to $\nu_\mu \to \nu_e$
oscillations, MiniBooNE is expected to detect an excess of several
hundred of $\nu_e$ events during its first year of operation, establishing 
the oscillation signal at $8\sigma$ to $10\sigma$ level (see the left panel 
of fig. \ref{future}). If this happens, the second detector will be
installed, with the goal to accurately measure the oscillation parameters. 
MiniBooNE will begin taking data in 2002.    

A number of long baseline accelerator neutrino experiments have been 
proposed to date. They are designed to independently test the oscillation
interpretation of the results of the atmospheric neutrino experiments,
accurately measure the oscillation parameters and to (possibly) identify
the oscillation channel. The first of these experiments, K2K (KEK to
Super-Kamiokande), started taking data in 1999. It has a baseline
of 250 km, average neutrino energy $\langle E\rangle\simeq 1.4$ GeV and is
looking for $\nu_\mu$ disappearance. K2K should be able to test practically 
the whole region of oscillation parameters allowed by the SK atmospheric
neutrino data except perhaps the lowest-$\Delta m^2$ part of it (see fig.
\ref{future}). The data collected from June 1999 to June 2000 ($2.6\times 
10^{19}$ protons on target) have been reported \cite{Boyd}. The observed 
number of fully contained $\mu$-like event is 27 with a background of less
than $10^{-3}$ events. This has to be compared with the expected number 
$40.3^{+4.7}_{-4.6}$ assuming no neutrino oscillations. The observed 
deficiency of $\mu$-like events disfavours the no-oscillation
hypothesis at $2\sigma$ level and supports the neutrino oscillation 
interpretation of the SK atmospheric neutrino data. However, more 
data have to be accumulated before definitive conclusions can be drawn.

Two other long baseline projects,
NuMI - MINOS (Fermilab to Soudan mine in the US) \cite{Woj} and CNGS 
(CERN to Gran Sasso in Europe) \cite{Rub}, each with the baseline of 730
km, will be sensitive to smaller values of $\Delta m^2$ and should be able 
to test the whole allowed region of SK (fig. \ref{future}). MINOS will
look for $\nu_\mu$ disappearance and spectrum distortions due to $\nu_\mu 
\to\nu_x$ oscillations. It may run in three different energy regimes -- high, 
medium and low energy ($\langle E\rangle\simeq$ 12, 6 and 3 GeV,
respectively). MINOS is scheduled to start taking data in 2003. 
CERN to Gran Sasso ($\langle E\rangle\simeq$ 17 GeV) will be an appearance 
experiment looking specifically for $\nu_\mu\to\nu_\tau$ oscillations. It 
will also probe $\nu_\mu$ disappearance and $\nu_\mu \to \nu_e$ appearance. 
At the moment, two detectors have been approved for participation in the   
experiment -- OPERA and ICARUS. The whole project was approved in December
of 1999 and the data taking is planned to begin in 2005 \cite{nuoscind}. 

Among widely discussed now future projects are neutrino factories -- muon
storage rings producing intense beams of high energy neutrinos.
In addition to high statistics studies of neutrino interactions,
experiments at neutrino factories should be capable of measuring neutrino
oscillation parameters with high precision and probing
the subdominant neutrino oscillation channels, matter effects and CP
violation effects in neutrino oscillations \cite{nufact}.

\section{Phenomenological neutrino mass matrices
\label{phen}}

As was discussed before, one needs at least four light neutrino species 
to accommodate the data of all neutrino experiments, which would imply the
existence of a light sterile neutrino. If, however, the LSND result, which 
has not yet been independently confirmed, is left out, one can describe 
all the data with just three usual neutrinos $\nu_e$, $\nu_\mu$ and
$\nu_\tau$. We shall discuss here 3-neutrino schemes and only briefly comment 
on the 4-neutrino schemes. 

As follows from the analyses of solar and atmospheric neutrino data, there
are two distinct mass squared difference scales in the three neutrino
framework, $\Delta m^2_{atm}\sim 10^{-3}\,\mbox{eV}^2$ and $\Delta m^2_\odot
\aprle 10^{-4}\,\mbox{eV}^2$. The hierarchy $\Delta m^2_{atm}\gg \Delta
m_\odot^2$ means that one of the three neutrino mass eigenstates (which we 
denote $\nu_3$) is separated by the larger mass gap from the other two.
One than has to identify $|\Delta m_{31}|^2\simeq |\Delta m_{32}|^2 = 
\Delta m_{atm}^2$, $|\Delta m_{21}|^2=\Delta m_\odot^2$. 

We know already from the 2-flavour analysis that $\nu_\mu \leftrightarrow 
\nu_\tau$ should be the main channel of oscillations whereas 
$\nu_e\leftrightarrow \nu_{\mu,\tau}$ oscillations can only be present as 
the subdominant channels. In the 3-flavour framework this means that the 
element $U_{e3}$ of the lepton mixing matrix $U_{ai}$ is small. This is
in accord with the CHOOZ limit (\ref{choozres}). The smallness of $U_{e3}$ 
means that the $\nu_\mu \leftrightarrow \nu_\tau$ and $\nu_e\leftrightarrow 
\nu_{\mu,\tau}$ oscillations 
approximately decouple, and the 2-flavour analysis gives a good first
approximation. In terms of the standard parametrization of the lepton mixing 
matrix, the mixing angle describing the main channel of the atmospheric 
neutrino oscillations is $\theta_{23}$, and its best-fit value following from 
the SK data is $45^\circ$. This fact and the smallness of $|U_{e3}| \equiv 
\sin\theta_{13}$ mean that the mass eigenstate $\nu_3$ mainly consists of 
the flavour eigenstates $\nu_\mu$ and $\nu_\tau$ with approximately 
equal weights, while the admixture of $\nu_e$ to this state is small or zero.
Together with the unitarity of the lepton mixing matrix, this implies  
that solar neutrino oscillations, which are governed by the mixing angle
$\theta_{12}$, transform the solar $\nu_e$ into a superposition of $\nu_\mu$ 
and $\nu_\tau$ with equal or almost equal weights. This holds irrespective
of whether the solution of the solar neutrino problem is LMA, SMA, LOW or VO. 

With $\Delta m_\odot^2 \ll \Delta m_{atm}$ there are three possible types 
of neutrino mass ordering. The first is the ``normal'', or direct hierarchy 
$m_1, m_2\ll m_3$; the second possibility is an inverted mass hierarchy 
$m_3\ll m_1\simeq m_2$. The present-day data
do not discriminate between the normal and inverted hierarchies; such a
discrimination may become possible in future if the earth's matter effects
in atmospheric or long baseline $\nu_e \leftrightarrow \nu_\mu$ or $\nu_e
\leftrightarrow \nu_\tau$ oscillations are observed. Finally, neutrinos 
may be 
quasi-degenerate in mass, with only their mass squared differences being
hierarchical. Direct neutrino mass measurements allow the average neutrino 
mass as large as a few eV (provided that the $2\beta 0\nu$ constraint
(\ref{meff}) is satisfied). In that case neutrinos could constitute a 
noticeable fraction of the dark matter of the universe (hot dark matter).

The experimental information on the neutrino masses and lepton mixing
angles allows one to reconstruct the phenomenologically allowed forms of
the neutrino mass matrix $M_\nu$. 
%
For example, assuming $\theta_{23}=\pi/4$ 
and the direct mass hierarchy, one finds (in the basis in which the mass 
matrix of charged leptons is diagonal) a simple zeroth order texture for 
$M_\nu$ which is the first matrix on the r.h.s. of eq. (\ref{M1}): 
\be
M_\nu \propto 
\left(\begin{array}{ccc}
0    &   0     & 0 \\
0    &   1     & 1 \\   
0    &   1     & 1 
\end{array}\right)\, \Rightarrow 
\left(\begin{array}{ccc}
\kappa      & \varepsilon       & \varepsilon' \\
\varepsilon & ~1+\delta-\delta' & 1-\delta \\   
\varepsilon' & ~1-\delta & 1+\delta+\delta'
\end{array}
\right).
\label{M1}
\ee
It yields $\Delta m_\odot^2\equiv \Delta m_{21}^2=0$, and so to get 
a realistic mass matrix one has to fill in zero entries with small nonzero 
terms and also to slightly perturb the large entries, i.e. to modify $M_\nu$  
as shown in eq. (\ref{M1}) (the parameters $\kappa$, $\varepsilon$, 
$\varepsilon'$ $\delta$ and $\delta'$ are assumed to be small). Similarly, one 
can find zeroth order textures for the inverted mass hierarchy: 
\be
M_\nu \propto 
\left(\begin{array}{ccc}
\pm 2    & ~~0     & ~~0 \\
~~0    &   ~~1     & -1 \\   
~~0    &   -1    & ~~1 
\end{array}\right); \quad 
\left(\begin{array}{ccc}
~~0     & ~~  1     & \pm 1\\
~~1     & ~~  0     &  ~~0   \\   
\pm1   &  ~~ 0     &  ~~0 
\end{array}\right), 
\label{M2}
\ee
from which realistic mass matrices can obtained through the 
\mbox{modifications} analogous to those in eq. (\ref{M1}). 
The small entries of the neutrino mass matrices have to satisfy certain 
constraints \cite{Akh2}.

The fact that at least one of the lepton mixing angles, $\theta_{23}$, is 
large came as a surprise to many people. The general expectations were 
that the mixing angles in the leptonic sector are small, as they are in 
the quark sector. The smallness of the quark mixing angles, i.e. the fact
that the CKM matrix $V=V_u^\dag V_d$ is close to the unit matrix, implies 
$V_u\simeq V_d$. This can be understood as an indication that the up-type 
and down-type quarks get their masses in a similar way, so that the
unitary matrices $V_u$ and $V_d$ of left-handed rotations that diagonalize 
the quark mass matrices $M_u$ and $M_d$ are similar. 

The lepton mixing matrix is also a product of two unitary matrices, the 
matrices of left-handed rotations of the mass matrices of charged leptons 
and neutrinos: $U=U_{l}^\dag U_\nu$. However, charged leptons and neutrinos 
are different in a very important way: the former are charged and can only be 
Dirac particles whereas the latter are neutral and can also be Majorana
particles. Majorana particles can get their masses in a way 
which is very different from that of Dirac particles. In particular,
neutrino masses can be generated by the seesaw mechanism or radiatively.
Thus, there are no {\em a priori} reasons to expect that $U_l\simeq U_\nu$ 
and that the lepton mixing is small. Actually, the fact that $\theta_{23}$
is large may be an indirect indication that neutrinos are Majorana
particles. 

We don't know yet if the mixing angle $\theta_{12}$ that governs the solar 
neutrino oscillations is large or small, though the current data seem
to favour large values of $\theta_{12}$. At the same time, we know that 
the mixing angle $\theta_{13}$ is relatively small (or may be very small), 
see eq. (\ref{choozres}). 
This mixing angle can be probed in future long baseline experiments which
are going to study $\nu_e\leftrightarrow \nu_{\mu,\tau}$ appearance, earth
matter effects on neutrino oscillations and CP violation in the leptonic 
sector. It can also be probed in reactor neutrino experiments and through
the detection of neutrinos from a future galactic supernova. 
%
CERN to Gran Sasso (and possibly Fermilab to Soudan mine) experiments are
expected to be able to measure the values of $|U_{e3}|^2$ of the order of 
$10^{-2}$ \cite{Rub}, reactor experiments could probe the values of
$|U_{e3}|^2$ down to $\sim 3\times 10^{-3}$ \cite{Mik}, supernova neutrino 
detection could have an order of magnitude better sensitivity \cite{DiSm}, 
and neutrino factories may be able to probe as small values of 
$|U_{e3}|^2$ as a few $\times 10^{-5}$ \cite{nufact}. 

Can we understand 
the fact that $|U_{e3}|=\sin\theta_{13}$ is small even though $\theta_{23}$
and probably also $\theta_{12}$ are large? And can we get at least a rough 
idea of how small $|U_{e3}|$ actually is? The latter would be of great
importance for 
future long baseline experiments. 

One can obtain an order of magnitude estimate of $|U_{e3}|$ if one assumes 
that there is no fine tuning between the elements (12) and (13) of the 
neutrino mass matrix $M_\nu$ (in the notation of eq. (\ref{M1}), between 
$\varepsilon$ and $\varepsilon'$). Then, neglecting possible leptonic CP 
violation effects, one finds the following approximate relation between 
$U_{e3}$, $\theta_{12}$, $\Delta m_\odot^2$ and $\Delta m_{atm}^2$ \cite{ABR}: 
\be
U_{e3}^2 \simeq \frac{1}{4}\cdot\frac{\tan^2 2\theta_{12}}{(1+\tan^2
2\theta_{12})^{1/2}}\cdot\frac{\Delta m_\odot^2}{\Delta m_{atm}^2}\,.
\label{Ue3}
\ee
This expressions provides an explanation of the smallness of $U_{e3}$: it 
is a consequence of the hierarchy $\Delta m_\odot^2\ll \Delta m_{atm}^2$. 
Since the value of $\Delta m_{atm}^2$ is fixed (within a factor of 3 or
so) by the SK atmospheric neutrino data, the actual value of $U_{e3}$ 
depends on the solution of the solar neutrino problem. In particular, for 
the LMA solution (which is currently the preferred one) eq. (\ref{Ue3}) 
predicts rather large values of $U_{e3}$ which may be just below the 
CHOOZ limit (\ref{choozres}). This means that $U_{e3}$ may be measured
soon! One should, however, keep in mind that eq. (\ref{Ue3}) gives only 
plausible values of $U_{e3}$ as its predictions are crucially based on the 
assumption of no fine tuning between the elements $\varepsilon$ and 
$\varepsilon'$ of the neutrino mass matrix (\ref{M1}). Such a fine tuning 
may, actually, be there and be natural if it is enforced by a flavour
symmetry.

Four-neutrino schemes can be analyzed similarly to the 3-neutrino 
case. The data allow essentially two 2+2 schemes. In each of these schemes
there are two pairs of nearly degenerate mass eigenstates separated 
by large $\Delta m_{LSND}^2$. The mass splittings between the components of 
the quasi-degenerate pairs are $\Delta m_\odot^2$ and $\Delta m_{atm}^2$.
There are also 3+1 schemes in which one of the mass eigenstates is
separated by the large $\Delta m_{LSND}^2$ from the rest 
of the states. The most attractive 3+1 scheme is the one in which the 
lone mass eigenstate is predominantly the sterile neutrino; in this 
scheme the 3-flavour dynamics of solar and atmospheric neutrino
oscillations is only slightly perturbed.  Until recently, this scheme has 
been considered experimentally disfavoured (compared to the 2+2 schemes),
but with the recent SK data the 
situation has changed. In 2+2 schemes the total contribution of 
sterile neutrinos to oscillations of solar and atmospheric neutrinos is equal 
to unity, so the fact that SK disfavours the pure active to sterile 
oscillations of {\em both} atmospheric and solar neutrinos makes the
2+2 schemes less plausible.  At the same time, the recently reported new 
analysis of the LSND data \cite{LSND} has shifted the allowed parameter
space towards smaller mixing angles, which improves the fit within the 3+1 
scheme. Therefore the 3+1 scheme in which the lone mass eigenstate is 
predominantly $\nu_s$ is now also acceptable 
\cite{Sm2,PS,Barger2}. Different 3+1 schemes can also fit the data \cite{GL}. 
Other recent discussions of the 4$\nu$ schemes can be found in  
\cite{Lisi1,Giu,4nu,C}).

The neutrino mass matrix textures can provide us with a hint of the 
symmetries or dynamics underlying the theory of neutrino mass. With the
forthcoming data from future neutrino experiments, it may eventually
become possible to uncover the mechanism of the neutrino mass generation,
which may hold the clue to the general fermion mass problem.

It is very likely that in a few years from now new experiments will bring
us the answers to many questions about neutrinos and finally allow us to solve 
the solar neutrino problem. Neutrinos may also bring us new surprises, as
they did many times in the past.

This work was supported by Funda\c{c}\~ao para a Ci\^encia e a Tecnologia
through the grant PRAXIS XXI/BCC/16414/98 and also in part by the TMR
network grant ERBFMRX-CT960090 of the European Union.

\end{document}